\documentclass{article}
\usepackage{graphics}
\usepackage{epsfig}

\begin{document}

\author{D. B. ION$^{1,2)}$ and M. L. ION$^{3)}$ \\
$^{1)}$ National Institute for Physics and Nuclear
Engineering \\ ''Horia Hulubei'', Bucharest, P.O.Box MG-6, Romania; \\
$^{2)}$ TH Division, CERN, CH-1211 Geneva 23, Switzerland; \\
$^{3)}$ University of Bucharest, Department of Atomic and Nuclear
Physics, \\ P.O.Box MG-11, Romania}

\title{{\bf Saturation of optimality limits in hadron-hadron scatterings}.}
\maketitle

\begin{abstract}
In this paper the optimal unitarity lower bound on logarithmic slope of the
diffraction peak is investigated. It is shown that the unitarity lower bound
is just the {\it optimal logarithmic slopes} predicted by the {\it %
principle of least distance in space of states. }A systematic
tendency towards the saturation of {\it the most forward-peaked
limit} is observed from all the available experimental data of all
principal hadron-hadron (e.g., $PP,$ $\overline{P}P,$ $K^{\pm }P,$
$\pi ^{\pm }P$ ) scatterings practically at all{\it \
}laboratory momenta{\it .}
\end{abstract}
\bigskip\ 
{\bf 1. Introduction}
\bigskip\ 

Recently, in Ref. [1], by using {\it reproducing kernel Hilbert space }%
(RKHS\} methods [2-4], we described the quantum scattering of the spinless
particles by a {\it principle of minimum distance } {\it in the space
of the scattering states (PMD-SS). }Some preliminary experimental tests of
the {\it PMD-SS, even }in the crude form [1] when the complications due
to the particle spins are neglected, showed that the actual experimental
data for the differential cross sections of all $PP,$ $\overline{P}P,$ $%
K^{\pm }P,$ $\pi ^{\pm }P,$ scatterings at all energies higher than 2 GeV,
can be well systematized by {\it PMD-SS }predictions{\it .} Moreover,
connections between the {\it optimal states [1], the PMD-SS in the space
of quantum states and }the {\it maximum entropy principle for the
statistics of the scattering channels }was also recently established by
introducing {\it quantum scattering entropies }[5-7]. Then, it was shown
that the experimental pion-nucleon as well as pion-nucleus scattering
entropies are well described by optimal entropies and that the experimental
data are consistent with the{\it \ principle of minimum distance in the
space of scattering states (PMD-SS) }[1].

An important model independent result obtained Ref. [1], via the{\it \ }%
description of quantum scattering by the{\it \
principle of minimum distance in space of states (PMD-SS)}, is the following
{\it optimal lower bound on logarithmic slope of the forward diffraction
peak} in hadron-hadron elastic scattering:

\begin{equation}  \label{1}
b\equiv \frac d{dt}\ln \left[ \frac{d\sigma }{d\Omega }(s,t)\right] _{\mid
t=0}\geq b_o\equiv \frac{\overline{\lambda }^2}4\left[ \frac{4\pi }{\sigma
_{el}}\frac{d\sigma }{d\Omega }(1)-1\right]
\end{equation}
where$\sqrt{s}$ and $\sqrt{\left| t\right| }$ are c.m. energy and transfer
momentum, respectively, while$\,\overline{\lambda }$ is c.m. d' Broglie wave
length, $\frac{d\sigma }{d\Omega }(x)$ is the differential cross section, $%
x\equiv \cos \theta ,$ being the c.m. scattering angle, and $\sigma _{el}$
is the integrated elastic cross section. So, the quantum scattering state
which saturate the optimal bound (1) is {\it the most forward-peaked
quantum state}. The optimal bound (1) improves in a more general and exact
form not only the unitarity bounds derived by Martin [8], MacDowell and
Martin [9] (see also Ref. [10]) for the logarithmic slope $b_{A}$ of
absorptive contribution $\frac{d\sigma _A}{d\Omega }(s,t)$ to the elastic
differential cross sections but also the unitarity lower bound derived in
Ref. [11] for the slope $b$ of the entire $\frac{d\sigma }{d\Omega }(s,t)$
differential cross section. Therefore, it would be very important to clarify
the {\it optimal status of the bound }(1) and to make an experimental
detailed investigation of its saturation in the hadron-hadron scattering
especially in the low energy region.

The aim of this paper is twofold: to show that the optimal bound (1) is the
''{\it singular}'' solution of the problem {\it to find an unitarity
lower bound on the logarithmic slope b} {\it when: }$\sigma _{el}${\it %
\ and }$\frac{d\sigma }{d\Omega }(1)$ {\it are given}, and, to obtain
experimental tests of the bound (1) at all available energies for the
principal hadron-hadron elastic scatterings: $\pi ^{\pm }P\rightarrow \pi
^{\pm }P,$ $K^{\pm }P\rightarrow K^{\pm }P,$ $PP\rightarrow PP,$ and $%
\overline{P}P\rightarrow \overline{P}P$. So, in Sect. 2 the
problem of the optimal status of the bound (1) is completely
solved by showing that this bound is the singular solution
($\lambda _0=0$) in the unconstrained Lagrangean minimization
problem. The relation of the optimal bound (1) with the
Martin-MacDowell bound as well as its saturations in different
particular cases (e. g. pure absorptive cases, etc.) are
discussed. Some details about the PMD-SS-history as well as about
other predictions of the principle of minimum distance in the
space of states are presented in Sect. 3. Then, it is shown that
the{\it \ optimal state} obtained with the simplified version
of the {\it PMD-SS }have not only the property that is
{\it the most forward-peaked quantum state }but also possesses
many other
peculiar properties such as: (i) scaling in the variable $\tau _o=2\sqrt{%
|t|b_o}$ , (ii) {\it maximum Tsallis-like scattering entropies }[6,7] $%
S_\theta (q),$ $S_L(q)$, $S_{\theta L} (q)$, $\tilde{S}_{\theta L}
(q)$ for the all positive nonextensivity parameters $q$, etc.,
that make it a good candidate for the description of the quantum
scattering. The results on the experimental test of the
{\it saturation of the most forward-peaked state limit} are
presented in Sect. 4 while the Sect. 5 is reserved for summary and
conclusions.
\bigskip\

{\bf2. The problem of the unitarity lower bound on logarithmic
slope}
\bigskip\ 

Let us start with the hadron-hadron elastic scattering process of spinless
particles for which the {\it description of the scattering amplitude }$%
f(x)${\it \ of the system} is given in terms of {\it partial amplitudes%
} $f_l,\,\,l=0,1,...$as

\begin{equation}  \label{2}
f(x)=\sum (2l+1)f_{l}P_{l}(x),\,\,\,\,\,\,\,\,\,x\in [-1,1],\,f_{l}\in C
\end{equation}
where $P_{l}(x),\,l=0,1,...$ , are Legendre polynomials, $f_{l},l=0,1,...$
being the partial amplitudes. If the normalization is chosen such that $%
\frac{d\sigma }{d\Omega }(x)=\mid f(x)\mid ^{2},$ {\it the elastic
integrated cross section }$\sigma _{el}$, is expressed in terms of partial
amplitudes by

\begin{equation}  \label{3}
\frac{\sigma _{el}}{2\pi }=2\sum (2l+1)\mid f_{l}\mid ^{2}=\Vert f\Vert ^{2}
\end{equation}
Now, a rigorous{\it \ unitarity lower bound} on the slope parameter b can
be obtained by solving completely the following {\it constrained
minimization} problem:

\begin{equation}  \label{4}
\min \{b\},\; when\mathit{\ }\sigma _{el}=fixed,\mathit{\ }\frac{d\sigma }{%
d\Omega }(1)=fixed,\: \mid f_l\mid ^2\leq Imf_l, \thinspace \thinspace
\thinspace l=0,1,..
\end{equation}
which is equivalently to the following {\it unconstrained minimization
problem}:
\begin{equation}  \label{5}
L\equiv \lambda _0b+ \lambda _1\left[ \frac{\sigma _{el}}{4\pi }-\sum
(2l+1)\mid f_l\mid ^2\right] +\lambda _2\left[ \frac{d\sigma }{d\Omega }%
(1)-\mid \sum (2l+1)f_l\mid ^2\right] +\sum \xi _lg_l\rightarrow \min
\end{equation}
where $g_l\equiv Imf_l-\mid f_l\mid ^2,$ $\lambda _i,\,\,i=0,1,2,$ and $\xi
_l$ , $l=0,1,...,$ are the corresponding Lagrange multipliers, and $b=(R%
\stackrel{\circ }{R}+I\stackrel{\circ }{I})/(R^2+I^2),$ where $R\equiv \sum
(2l+1)Ref_l,$ $I\equiv \sum (2l+1)Imf_l$, and $\stackrel{\circ }{R}\equiv
\sum (2l+1)\frac{l(l+1)}2Ref_l$, $\stackrel{\circ }{I}\equiv \sum (2l+1)
\frac{l(l+1)}2Imf_l$.

The non singular solution $(\lambda _0\neq 0)$ of the minimization problem
(5) was completely treated in Ref. [11]. Here we must underline that,
according to the rules from Ref. [12], the singular solution $\lambda _0=0,$
if it exists, can be essential in obtaining a rigorous unitarity lower bound
on the logarithmic slope parameter $b$. In order to prove this statement,
let us consider as an simple example the minimization problem: $x\rightarrow
\min $, subject to $x^2+y^2=0$. The solution of this problem is evident: $%
x_{\min }=y_{\min }=0$. Now, let use the Lagrange multiplier's problem with $%
\lambda _0\neq 0$: $L=x+\lambda (x^2+y^2).$ Then, we obtain the following
incompatible equation: $1+2\lambda x=0$, $2\lambda y=0$, and $x^2+y^2=0$.
Hence, we must consider the case $\lambda _0=0$ from which we get the
correct solution. Such a singular solution of the Lagrange function (5) can
be obtained by solving the {\it minimization problem}
\begin{equation}  \label{6}
\left\{ \sum (2l+1)\mid f_l\mid ^2+\alpha \left[ \frac{d\sigma }{d\Omega }%
(1)-\mid \sum (2l+1)f_lP_l(1)\mid ^2\right] \right\} \rightarrow \min
\end{equation}
The problem (6) is just the problem of {\it minimum norm }({\it %
minimum distance in the space of states}) when $\frac{d\sigma }{d\Omega }(1)$
is fixed. Thus, the unique solution of (6) exists and is given the following
optimal state (see Ref. [1])

\begin{equation}  \label{7}
f_{o1}(x)=f(1)\frac{K(x,1)}{K(1,1)}=f(1)\frac{\stackrel{.}{P}_{L_o+1}(x)+%
\stackrel{.}{P}_{L_o}(x)}{\left( L_o+1\right) ^2}
\end{equation}
where $\stackrel{.}{P}_l(x)\equiv dP_l(x)/dx\,\,$while the{\it \
reproducing kernel} $K(x,1)$ is given as follows

\smallskip\

$K(x,1)=\frac 12\sum_{l=0}^{L_o}(2l+1)P_l(x)=\frac 12\{\stackrel{.}{P}%
_{L_o+1}(x)+\stackrel{.}{P}_{L_o}(x)\}$,$\,\,\,K(1,1)=\left( L_o+1\right)
^2/2$,

\smallskip\

and

\begin{equation}  \label{8}
L_o=integer\left\{ \left[ \frac{4\pi }{\sigma _{el}}\frac{d\sigma }{d\Omega }%
(1)\right] ^{1/2}-1\right\}
\end{equation}
Hence, the singular solution ($\lambda _0=0,\,\,\chi _l=0,\,\,l=0,1,....$)
of the minimization problem (5) (or (4)) is just the {\it optimal slope}
\begin{equation}  \label{9}
b_o=\frac d{dt}\ln \left[ \mid f_{o1}(x)\mid ^2\right] _{\mid t=0}=
\overline{\lambda }^2\frac{L_o(L_o+2)}4=\frac{\overline{\lambda }^2}4\left[
\frac{4\pi }{\sigma _{el}}\frac{d\sigma }{d\Omega }(1)-1\right]
\end{equation}
This result is in agreement with that obtained in Ref. [1] directly from the
fundamental inequality $\mid f(y)\mid ^2\leq \frac{\sigma _{el}}{2\pi }%
K(y,y),$ $y\in \left[ -1,+1\right] .$ Then, by developing the scattering
amplitude $f(y)$ and the {\it reproducing kernel function }$K(y,y)$ in
Taylor's series around the point $y=1$ and $L=L_o$, it was shown that
\begin{equation}  \label{10}
\frac d{dy}\left[ \frac{d\sigma }{d\Omega }(y)\right] _{\mid \;y=1}\geq
\frac{\sigma _{el}}{2\pi }\cdot \frac d{dy}\left[ K(y,y)\right] _{\mid \;
y=1,L=L_o}=\frac{d\sigma }{d\Omega }(1)\frac{L_o(L_o+2)}2
\end{equation}
From (10) and (8) we get the optimal lower bound (1) since, according to the
definition of the logarithmic slope $b$ combined with the kinematical
relations $t=-2q^2(1-y)$, we have $dt=2dy/\overline{\lambda }^2$ and
\begin{equation}  \label{11}
b=\frac{\overline{\lambda }^2}2\frac d{dy}\left[ \frac{d\sigma }{d\Omega }%
(y)\right] _{\mid \; y=1}/\frac{d\sigma }{d\Omega }(1)
\end{equation}

Now, by comparing the singular solution (1) with that $b_{\min }(\lambda
_0\neq 0)$ obtained in Ref. [11] for the non singular case $\lambda _0\neq
0, $ we see that: $b_o>b_{\min }(\lambda _0\neq 0)\simeq \frac 89b_o$.
Consequently, since we get the chain of bounds: $b\geq b_o>b_{\min }(\lambda
_0\neq 0),$ the unique solution of the problem (5) is that given by the
optimal bound (1) since max$\{b_{\min }(\lambda _0=0),b_{\min }(\lambda
_0\neq 0)\}=b_o$

{\it Remark 1}: The solution to the minimization problem (5) is
of relevance since in the pure absorptive case ($Ref(x)=0$) the singular
solution $b\geq b_o$ as well as that non singular $b>b_{\min }(\lambda
_0\neq 0)\simeq \frac 89b_o$ (see Ref.[11]) are going in

\begin{equation}  \label{12}
b_{A}\geq \frac{\overline{\lambda }^{2}}{4}\left[ \frac{\sigma _{T}^{2}}{%
4\pi \overline{\lambda }^{2}\sigma _{el}}-1\right] \equiv b_{o}^{A}
\end{equation}
and, respectively, in the MacDowell-Martin bound
\begin{equation}  \label{13}
b_{A}>\frac{8}{9}b_{o}^{A}=\frac{2\overline{\lambda }^{2}}{9}\left[ \frac{%
\sigma _{T}^{2}}{4\pi \overline{\lambda }^{2}\sigma _{el}}-1\right] \equiv
b_{MD-M}
\end{equation}
since
\begin{equation}  \label{14}
\frac{d\sigma ^{A}}{d\Omega }(1)=[\mathop{\rm Im}f(x)]^{2}=\sigma
_{T}^{2}/(4\pi \overline{\lambda })^{2},\smallskip\ \mathrm{an dd }b=b_{A}
\end{equation}

{\it Remark 2}: Let $\sigma _{el}$ and $\sigma _{T}$ be fixed
from experiment let us consider the minimization problem

\begin{equation}  \label{15}
\min \{b\},\mathrm{when} \sigma _{el}=fixed,\mathrm{and} \sigma
_{T}=fixed,\; \mid f_{l}\mid ^{2}\leq Imf_{l},\; l=0,1,.
\end{equation}
which is equivalent to the minimization of the Lagrangean
\begin{equation}  \label{16}
L\equiv \lambda _{0}b+\lambda _{1}\left[ \frac{\sigma _{el} }{4\pi }-\sum
(2l+1)\mid f_{l}\mid ^{2}\right] +\lambda _{2}\left[ \sigma _{T}-\sum
(2l+1)Imf_{l}\right] +\sum \xi _{l}g_{l}\rightarrow \min
\end{equation}
Then, the unique solution of Eq. (15) is given by
\begin{equation}  \label{17}
b\geq \max \{b_{\min }(\lambda _{0}=0),b_{\min }(\lambda _{0}\neq 0)\}=\max
\{b_{o}^{A},b_{MD-M}\}=b_{o}^{A}
\end{equation}

The quantum scattering state which saturate the optimal bound (23) is
{\it the most forward-peaked quantum state }which is obtained as the
absorptive limit of the optimal state (7).

Indeed, the singular solution $\lambda _{0}=0$ of the problem (16), can be
obtained by solving the {\it minimization problem}
\begin{equation}
\left\{ \sum (2l+1)\mid f_{l}\mid ^{2}+\alpha \left[ \sigma _{T}-\sum
(2l+1)Imf_{l}\right] \right\} \rightarrow \min  \label{18}
\end{equation}
The unique solution of this problem is given by
\begin{equation}
\begin{array}{c}
Ref_{l}^{o}=0,\smallskip \ \mathrm{for all }l, \\
Imf_{l}^{o}=4\pi \overline{\lambda }\cdot \frac{\sigma _{el}}{\sigma _{T}},%
\mathrm{for }\; 0\leq l\leq L_{o}^{A},\smallskip \ Imf_{l}=0, \mathrm{for }%
l\geq L_{o}^{A}+1
\end{array}
\label{19}
\end{equation}
where
\begin{equation}
L_{o}^{A}=\mathrm{integer}\left\{ \left[ \frac{\sigma _{T}^{2}}{4\pi
\overline{\lambda }^{2}\sigma _{el}}\right] ^{1/2}-1\right\} =\mathrm{integer%
}\left\{ \left[ \frac{4\pi }{\sigma _{el}}\frac{d\sigma }{d\Omega }%
^{A}(1)\right] ^{1/2}-1\right\}  \label{20}
\end{equation}
By introducing the results (19) and (20) in Eq. (2) we obtain
\begin{equation}
\begin{array}{c}
Ref_{o1}^{A}(x)=0 \\
Imf_{o1}^{A}(x)=Imf(1)\frac{\sum (2l+1)P_{l}(x)}{(L_{o}^{A}+1)^{2}}=Imf(1)%
\frac{\stackrel{.}{P}_{L_{o}+1}(x)+\stackrel{.}{P}_{L_{oT}}(x)}{\left(
L_{o}^{A}+1\right) ^{2}}
\end{array}
\label{21}
\end{equation}

Hence, for purely absorptive amplitudes the optimal solution to the
minimization problem (19) exists and is given by Eq. (21) which in fact is
the pure absorptive limit of the optimal state (7) but with $L_o=L_o^A.$
Hence, in the pure absorptive case from (20) we get
\begin{equation}  \label{22}
b_{\min }(\lambda =0)=\overline{\lambda }^2\frac{L_o^A(L_o^A+2)}4= \frac{%
\overline{\lambda }^2}4\left[ \frac{\sigma _T^2}{4\pi \overline{\lambda }%
^2\sigma _{el}}-1\right] \equiv b_o^A
\end{equation}

Therefore, the unique solution of the problem (15) is given by

\begin{equation}  \label{23}
b\geq \max \{b_{o}^{A},b_{MD-M}^{A}\}=b_{o}^{A}
\end{equation}

The MacDowell-Martin bound $b^A>b_{MD-M}^A$ (13) emerge as the non singular
solution in a special case of the minimization problem (5) when $Ref(x)=0$
and due the fact that is $\frac 89b_o^A$ is eliminated in the final solution
(23) of the problem (15). Hence, the bound (1) include in a more general and
exact form the MacDowell-Martin bound since the validity of the bound $b\geq
b_o$ includes the validity of any inequality obtained as a consequence of
the inequality $b_o>a$, e.g., $a\equiv b_{MD-M}^A$(13).

Finally, we note that the fundamental inequality (16) from Ref. [1] tell us
that if $\sigma _{el}$ and$\frac{d\sigma }{d\Omega }(1)$ (or if $\sigma
_{el} $ and $\sigma _T$) are fixed from experiment, then the number $(L+1)$
of partial amplitudes in any phase shift analysis (PSA) must obey the
optimal bounds
\begin{equation}  \label{24}
\left( L+1\right) ^2\geq \frac{4\pi }{\sigma _{el}}\frac{d\sigma }{d\Omega }%
(1),\smallskip \ \left( L+1\right) ^2\geq \frac{\sigma _T^2}{4\pi \overline{%
\lambda }^2\sigma _{el}}
\end{equation}
respectively, or equivalently $L\geq L_o\geq L_o^A.$

\bigskip\
{\bf 3. Why principle of minimum distance in space of states?}
\bigskip\ 

As we already underlined in introduction, in Ref. [1] are investigated the
essential features of the hadron-hadron scattering by using the optimization
theory [1] in the Hilbert space of the partial scattering amplitudes. Then,
knowledge about the hadron-hadron scattering system (or more concretely,
about partial amplitudes) are deduced by assuming that it behaves as to
optimize some given{\it \ measure of its effectiveness}, and thus the
{\it behavior of the system} is completely specified by identifying the
criterion of effectiveness and applying constrained optimization to it. This
approach is in fact known as describing the system in terms of an {\it %
optimum principle.}

The earliest {\it optimum principle} was proposed by Hero of Alexandria
(125 B.C.) in his {\it Catoptrics} in connection with the behavior of
light. Thus, Hero of Alexandria (125 B.C.) proved mathematically the
following first genuine scientific{\it \ minimum principle} of physics.

Hero's principle of minimum distance (PMD) tell us that when a ray of light
is reflected by a mirror, the path actually taken from the object to the
observer`s eye is the shortest path from all possible paths. If we extend
the PMD-idea to the behavior of light in gravitational fields, then, we
obtain immediately that according to the PMD-optimum principle, modified to
include the interaction of light with the gravitational field, {\it the
light must move on a specific shortest path which is the geodesic. }Having
in mind this successful result, the {\it principle of minimum distance}
was recently [15] extended to the quantum physics by choosing the ''{\it %
partial transition amplitudes''} as fundamental physical
quantities since they are labelled with good quantum numbers such
as charge, angular momentum, isospin, etc. These physical
quantities are chosen{\it \ }as the{\it \ system variational
variables } while{\it \ }the {\it distance in the Hilbert
space of the quantum states} is taken as measure of the system
effectiveness expressed in terms of the system variables. The {\it %
principle of minimum distance in the space of states} (PMD-SS) is chosen as
variational optimum principle by which one should obtain those values of the
''{\it partial amplitudes}'' yielding minimum {\it effectiveness. }Of
course this new optimum principle{\it \ }can be formulated in a more
general mathematical form by using the S-matrix theory of the strong
interacting systems.

The Hilbert space {\bf H} of the scattering states, with the inner
product $<.$$\,,.>\,$ and the norm $\left\| \cdot \right\| $, defined on the
interval $S\equiv (-1,+1)$, can be given by

\begin{equation}  \label{25}
<f,g>=\int_{-1}^{+1}f(x)\overline{g(x)}dx\,,\,\,\,\,\,\,\,f,g\in \mathbf{H,}
\end{equation}

\begin{equation}  \label{26}
\left\| \,\,f\,\right\| ^2=<f,f>=\int_{-1}^{+1}\mid f(x)\mid ^2dx,\;\;f\in
\mathbf{H}
\end{equation}

If the {\it quantum distance} $D(f,g)$ between two quantum scattering
states $f$ and $g$ is defined [1, 13] by
\begin{equation}  \label{27}
D(f,g)=\min _\Phi \Vert f-g\exp (-i\Phi )\Vert =[\Vert f\Vert ^2+\Vert
g\Vert ^2-2\mid <f,g>\mid ]^{\frac 12}
\end{equation}
then the PMD-SS can be formulated in the following form.

PMD-SS: \quad {\it If $h$\ is the quantum state of the system when the
interaction is missing, then the true interacting quantum state $f$ of the
interacting system is that state which have the shortest distance $D(f,h)$\
in the space of interacting states compatible with the constraints imposed
by the interaction. (Of course, if $h=0$, then }$D(f,0)=\left\| f\right\| $.%
{\it )}

Therefore, let $M_{eff}\,$ be the measure of the scattering effectiveness
and let us consider the problems of optimization of few $M_{eff}$ defined as
follows:

(a) $M_{eff}\equiv D^{2}(f,0),$ or equivalently $M_{eff}=\sigma _{el}/4\pi $%
, where $\sigma _{el}${\it \ }is given in terms of partial amplitudes by
Eq. (3);

(b) $M_{eff}\equiv b$, where the {\it logarithmic slope }$b${\it \ of
the diffraction peak} is expressed in terms of the partial amplitudes as in

Sect. II;

(c) $M_{eff}\equiv S_{L}=-\sum (2l+1)p_{l}\ln p_{l},$ where: $p_{l}=4\pi
|f_{l}|^{2}/\sigma _{el},$ and $\sum (2l+1)p_{l}=1;$

(d)$\,\,M_{eff}\equiv S_{\theta }$ $=-\int\nolimits_{-1}^{+1}dx\,P(x)\,\ln
\,P(x)$,$\,$where $P(x)=\frac{2\pi }{\sigma _{el}}\frac{d\sigma }{d\Omega }%
(x)$, with $\int\nolimits_{-1}^{+1}dx\,P(x)=1;$

Now, let us consider the constrained optimization problems:
\begin{equation}  \label{28}
\mathrm{extremum}\{M_{eff}\}\;\mathrm{when}\;\sigma _{el} \;\mathrm{and}\;%
\frac{d\sigma }{d\Omega }(1)\;\mathrm{are\;fixed}.
\end{equation}
which, via Lagrange multipliers method [12], are reduced to the problems:

\begin{equation}  \label{29}
\pounds _F\equiv \lambda _0M_{eff}+\lambda _1\left[ \frac{\sigma _{el}}{4\pi
}-\sum (2l+1)\mid f_l\mid ^2\right] +\lambda _2\left[ \frac{d\sigma }{%
d\Omega }(1)-\mid \sum (2l+1)f_l\mid ^2\right] \rightarrow \mathrm{extremum}
\end{equation}
Hence, according to the general Lagrange multiplier rules [12], for all
these optimization problems (28)-(29) we are obligated to analyze in detail
the {\it singular solution} $\lambda _0=0$. Such a singular solution of
the Lagrangean function (29) can be obtained by solving the minimization
problem (6) which is just the{\it \ problem of minimum norm }(minimum
distance in the space of states)

\begin{equation}  \label{30}
\min [D(f,0)]\,\mathrm{when\ }\;\frac{d\sigma }{d\Omega }(1)\,\mathrm{\; is
\; fixed}\ fixed
\end{equation}
Thus, the unique solution of the problem (30) exists and is given by the
{\it PMD-SS-optimal state }(7) (see details in Ref. [1]). Therefore, the
{\it optimal effectiveness} are as follows (see e.g. Ref. [7] for the
(c), (d) cases):
\begin{equation}  \label{31}
\begin{array}{c}
M_{eff}^{o1}=b_o,\,\mathrm{in \;case (b)} \\
M_{eff}^{o1}=S_L^{o1}=\ln [2K(1,1)=\ln [(L_o+1)^2],\, \mathrm{in \;case \;(c)%
} \\
M_{eff}^{o1}=S_\theta ^{o1}=-\int\nolimits_{-1}^{+1}dx\left[ \frac{[K(x,1)]^2%
}{K(1,1)}\right] \ln \left[ \frac{[K(x,1)]^2}{K(1,1)}\right] ,\,\mathrm{in
\;case \;(d)}
\end{array}
\end{equation}
while the {\it optimal} {\it angular momentum }is given by Eq. (8).

Now, we recall briefly some peculiar properties of the PMD-SS-optimal state
(7) obtained in Refs. [1], [4]-[7] that can illustrate the importance of the
PMD-SS for an unified description of all quantum phenomena. These
characteristic features are as follows:

I. The {\it PMD-SS-optimal scaling} property (see the experimental tests
in Fig. 1 in Ref.[1])
\begin{equation}  \label{32}
\frac 1{\frac{d\sigma }{d\Omega }(s,1)}\,\frac{d\sigma ^{o1}}{d\Omega }%
(s,x)=\left[ \frac{K(x,1)}{K(1,1)}\right] ^2\approx \left[ \frac{2\,J_1(\tau
_o)}{\tau _o}\right] ^2,\mathrm{\,for\,\,small\,\,\,}\tau _o
\end{equation}
as a function of the {\it PMD-SS-optimal scaling variable} $\tau _o$%
\begin{equation}  \label{33}
\tau _o\equiv 2\,\left[ \left| t\right| b_{o1}\right] ^{1/2}=\left\{
\overline{\lambda }^2\,\left| t\right| \left[ \frac{4\pi }{\sigma _{el}}%
\frac{d\sigma }{d\Omega }(1)-1\right] \right\} ^{1/2}
\end{equation}

II. The {\it PMD-SS-optimal state} (7) is the {\it most forward peaked
scattering state} in the sense given by inequality (1) (see Figs. 1-3 in
this paper);

III. The {\it PMD-SS-optimal state }(7) is the state with {\it maximum
scattering entropies }(see proofs and the experimental illustrations in Ref.
[7]) in the sense that the entropies $\,\,S_L\,$and $S_\theta $ of any
scattering state must fulfill the inequalities
\begin{equation}  \label{34}
S_\theta \leq S_\theta ^{o1},\,\,\,S_L\leq S_L^{o1}
\end{equation}

IV. {\it PMD-SS-optimal state} (7) is the state of {\it equilibrium}
of the angular momenta channels considered as a {\it quantum statistical
ensemble. } Indeed, using Eqs. (31) and (34) we observe that the entropy $S_L
$ is similar to the Boltzmann entropy with a maximum value given by the%
{\it \ logarithm of number of the optimal states: }$S_L\leq S_L^{o1}=\ln
[(L_o+1)^2].\,$The results like (34) are also proved in Ref. [7] for the
nonextensive statistics [14] of the angular momenta channels.

V. From mathematical point of view, the {\it PMD-SS}-{\it optimal
states, }are functions of {\it minimum constrained} {\it norm} and
consequently can be completely described by{\it \ reproducing kernel
functions (}see Ref. [1,3-4]. {\it \ }So, with{\it \ }this respect the
{\it PMD}-SS-{\it optimal states} from the {\it reproducing kernel
Hilbert space (RKHS) }of the scattering amplitudes are analogous to the
{\it coherent} {\it states} from the RKHS of the {\it wave
functions.}

VI. For the scattering of particle with spins, {\it \ }the {\it %
PMD-SS-optimal states }are{\it \ s-channel helicity conserving} states
(see Refs. [4]). {\it \ }

Therefore, the{\it \ PMD-SS-optimal state} (7) have not only
the property that is {\it the most forward-peaked quantum state
}but also possesses many other peculiar properties such as I-V
that make it a good candidate for the description of the quantum
scattering via an optimum principle. So, the systematic
experimental investigation of the {\it optimal bound} (1) as
well as of the bounds (34) can be of great interest since, as a
direct signature of the {\it PMD-SS-optimal state dominance }in
the hadron-hadron scattering phenomena, the s{\it saturation of
the most
forward-peaked limit} as well as the {\it saturation of the} {\it %
maximum entropy limits }are expected to be experimentally
observed.

\bigskip\
{\bf 4. Experimental tests}
\bigskip\ 

Now, in order to obtain a consistent experimental test of the optimal
unitarity lower bound (1) we have compiled from the literature [7]-[18] the
experimental data on the diffraction slope parameter $b$ for $\pi ^{\pm
}P\rightarrow \pi ^{\pm }P,$ $K^{\pm }P\rightarrow K^{\pm }P,$ $%
PP\rightarrow PP,$ and $\overline{P}P\rightarrow \overline{P}P$ at all
available energies.

$\pi ^{\pm }P$-{\it scattering}: In this case, for $P_{LAB}\geq 3GeV/c$ ,
the experimental data on b, $\frac{d\sigma }{d\Omega }(1)$ and $\sigma _{el}$
are collected mainly from the original fit and data of Refs.[15]-[19]. To
these data we added some values of $b$ from the linear fit of Lasinski et
al. [24] and also calculated directly from the {\it phase shifts analysis}
(PSA) of Holer et al. [20]. In the low laboratory momenta $P_{LAB}\leq 2$
GeV/c, the slope parameters $b$ as well as the optimal slope $b_o$ are
determined in pairs $(b,b_o)$, at each laboratory momenta, from the PSA
[20]. Unfortunately, the values of $b_o$ corresponding to the Lasinski's
data [24] was impossible to be calculated since the values of $\frac{d\sigma
}{d\Omega }(1)$ from their original fit are not given.

$K^{\pm }P-${\it scattering. }The experimental data on b, $\frac{d\sigma
}{d\Omega }(1)$ and $\sigma _{el}$ , in the case of $K^{-}P$, are collected
from the original fit and data of Refs. [15], [17], [19], [21]. For $K^{+}P$%
-scattering, to these data from Refs. [15], [17], [19], [22] we added some
values of $b$ from the linear fit of Lasinski et al. [24] and also those
pairs $(b,b_o)$ calculated directly from the {\it experimental }(PSA)
solutions of Arndt et al. [23].

$\overline{P}P$ and $PP-${\it scatterings. }The experimental data on b, $%
\frac{d\sigma }{d\Omega }(1)$ and $\sigma _{el}$, in these cases, are
obtained from Refs. [19]-[21]. The data from original fits where $\frac{%
d\sigma }{d\Omega }(1)$ is given numerically was first selected for to
obtain the pairs $(b,b_o)$.

Now, for comparison of the experimental values of the slope parameter b with
the {\it optimal PMD-SS-optimal slope} $b_o$ we used a $(b,b_o)$-plot of
the elastic diffraction slopes as in Figs. 1a,b, where we plotted a total
number $N_p=420$ of experimental pairs $(b,b_o)$. From these pairs a number $%
N_p=125 $ pairs shown in Fig. 1b are coming from the experiments at $%
P_{LAB}\geq 2$ GeV/c. The dependence of the experimental slopes b and of the
{\it optimal PMD-SS-slope} $b_o$ (see Eq. (9)) on laboratory momenta ($%
P_{LAB}$) is presented in Figs. 2-3. The values of the $\chi
^2=\sum_j(b_j-b_{oj})^2/(\epsilon _{bj}^2+\epsilon _{b_oj}^2)$, (where $%
\epsilon _{bj}$ and $\epsilon _{b_oj}$ are the experimental errors
corresponding to $b$ and $b_o$ respectively)\ are used for the estimation of
departure from the {\it optimal PMD-SS-slope} $b_o$, and then, we obtain
the statistical parameters presented in Table 1.

\bigskip\
{\bf 5. Summary and Conclusion}
\bigskip\ 

The main results obtained in this paper can be summarized as follows:

(i) In this paper we proved that the optimal bound (1) is the singular
solution ($\lambda _0=0)$ of the problem {\it to find the unitarity lower
bound on the logarithmic slope b} {\it with the constraints imposed by
unitarity when }$\sigma _{el}${\it \ and }$\frac{d\sigma }{d\Omega }(1)$%
{\it \ are fixed. }This result is similar with that obtained recently in
Ref. [7] for the problem{\it \ to find an upper} {\it bound for the
scattering entropies when} $\sigma _{el}${\it \ and }$\frac{d\sigma }{%
d\Omega }(1)${\it \ are fixed from the experimental data.\thinspace }So,
the {\it PMD-SS-optimal state} is not only the most peaked state but also
a state which saturates the maximum entropy limit;

(ii) The optimal bound (1) is verified experimentally with high accuracy
(see the region b$\geq $b$_o$ in Fig. 1a) at all available energies for all
the principal hadron-hadron scatterings.

(iii) A systematic tendency toward the saturation of the{\it \ most
forward-peaked optimal state limit} is observed from the experimental data
on the logarithmic slope of all the $PP\rightarrow PP,$ $\overline{P}%
P\rightarrow \overline{P}P$, $\pi ^{\pm }P\rightarrow \pi ^{\pm }P$ and $%
K^{\pm }P\rightarrow K^{\pm }P$ scatterings at all available laboratory
momenta. The {\it proton-proton} ($\stackrel{\_}{\chi ^2}=5.06)$ and
{\it antiproton-proton } ($\stackrel{\_}{\chi ^2}=1.86)$ and $K^{+}P$ $(%
\stackrel{\_}{\chi ^2}=3.05)$ scatterings are examples where this saturation
of the most peaked optimal state limit is observed at all available
laboratory momenta. In fact the{\it \ }validity{\it \ of the principle
of least distance in space of states in hadron-hadron scattering} for p$%
_{LAB}\geq 2$ GeV/c is well illustrated in Fig. 1b and Table 1. A similar
tendency towards the saturation of the{\it \ most forward peaked
PMD-SS-optimal state limit}\ is observed (see Figs. 2-3) even in the low
energy regions at the energies between the resonances positions or/and in
the region corresponding to the ''{\it diffractive resonances} ''.

(iv) In order to see why the experimental logarithmic slopes b$_{\exp }($ as
well as the experimental entropies S$_L$ and S$_\theta $) are well described
by the {\it PMD-SS-optimal state (7)} we observe that the entropy $S_L$
is similar to the Boltzmann entropy with a maximum value given by the{\it %
\ logarithm of number of the optimal states}. Indeed, from (31) and (34) we
can conclude $(S_L\leq S_L^{o1}=\ln [(L_o+1)^2])$ that the {\it %
PMD-SS-optimal state} is the state of {\it equilibrium} of the angular
momenta channels considered as a {\it quantum statistical ensemble.}

One of us (D.B.I) would like to thank Prof. G. Altarelli for
fruitful discussions  as well as for hospitality during his stay in
TH-Division at CERN

\newpage\

\begin{description}
\item  {\bf Table 1} : $\stackrel{\_}{\chi ^{2}}-$statistical parameters
of the principal hadron-hadron scattering. In these estimations for P$%
_{LAB}\leq 2$ GeV/c the errors $\epsilon _{b_{o}}^{PSA}(\pi ^{\pm }P)=0.5$ $%
GeV^{-2}$ and $\epsilon _{b_{o}}^{PSA}(K^{+}P)=0.3$ $GeV^{-2}$ are taken
into account for the optimal slopes b$_{o}$ calculated from phase shifts
analysis [20] and [23], respectively.
\end{description}

\begin{center}
\begin{tabular}{lllll}
&  & N$_{p}($P$_{LAB}\geq 2GeV/c)$ &  & N$_{p}($all P$_{LAB})$ \\
Statistical parameters & N$_{p}$ & $\quad \,\quad \quad \chi ^{2}/$N$_{p}$ &
N$_{p}$ & $\quad \quad \chi ^{2}/$N$_{p}$ \\
$\quad P$ $P\rightarrow P$ $P$ & 29 & \quad $\quad \quad $5.01 & 32 & $\quad
\quad $5.06 \\
$\quad \stackrel{\_}{P}P\rightarrow \stackrel{\_}{P}P$ & 27 & \quad $\quad
\quad $0.56 & 45 & $\quad \quad $1.86 \\
$\quad \pi ^{+}P\rightarrow \pi ^{+}P$ & 12 & \quad $\quad \quad $2.90 & 43
& $\quad \quad $9.70 \\
$\quad \pi ^{-}P\rightarrow \pi ^{-}P$ & 14 & \quad $\quad \quad $2.61 & 68
& \quad $\quad $14.9 \\
$\quad K^{+}P\rightarrow K^{+}P$ & 20 & \quad $\quad \quad $0.75 & 55 & $%
\quad \quad $3.05 \\
$\quad K^{-}P\rightarrow K^{-}P$ & 23 & \quad $\quad \quad $2.18 & 97 & $%
\quad \quad $6.40 \\
$\quad $All experiments & 125 & \quad $\quad \quad $2.38 & 340 & \quad $%
\quad $6.83
\end{tabular}
\end{center}

\newpage
{\par\centering \epsfig{file=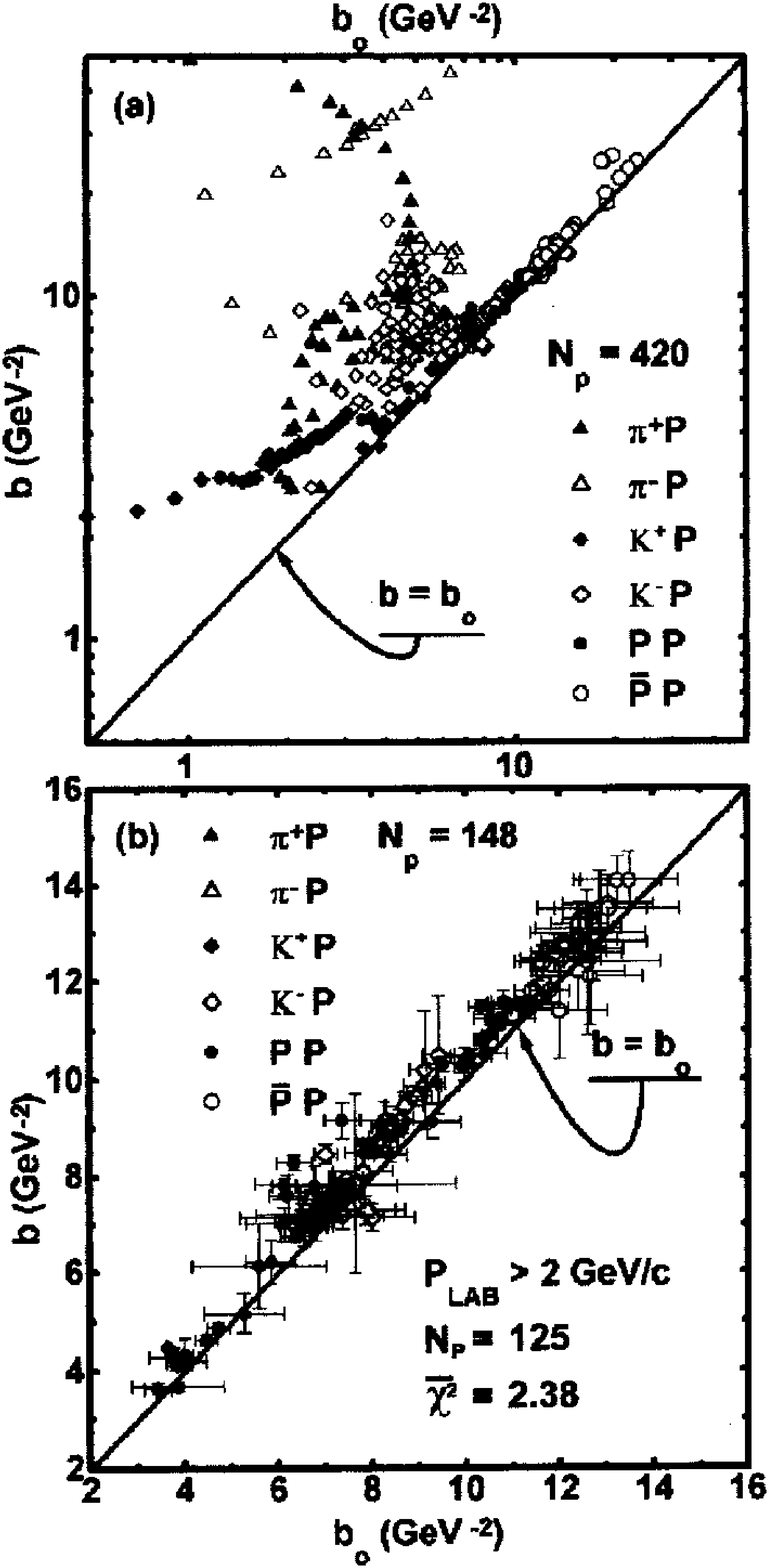, width=9.cm, height=17.5cm}}\\
Fig. 1:  The experimental values $b$ versus optimal values $b_{o}$ of
the logarithmic slope for all principal hadron-hadron scatterings: (a) at
all available momenta, and (b) only for $P_{LAB}\geq 2$ GeV/c. The
experimental data for $b$,$\frac{d\sigma }{d\Omega }(1)$ and $\sigma
_{el}$, are taken from Refs. [15]-[26].

\newpage
{\par\centering \epsfig{file=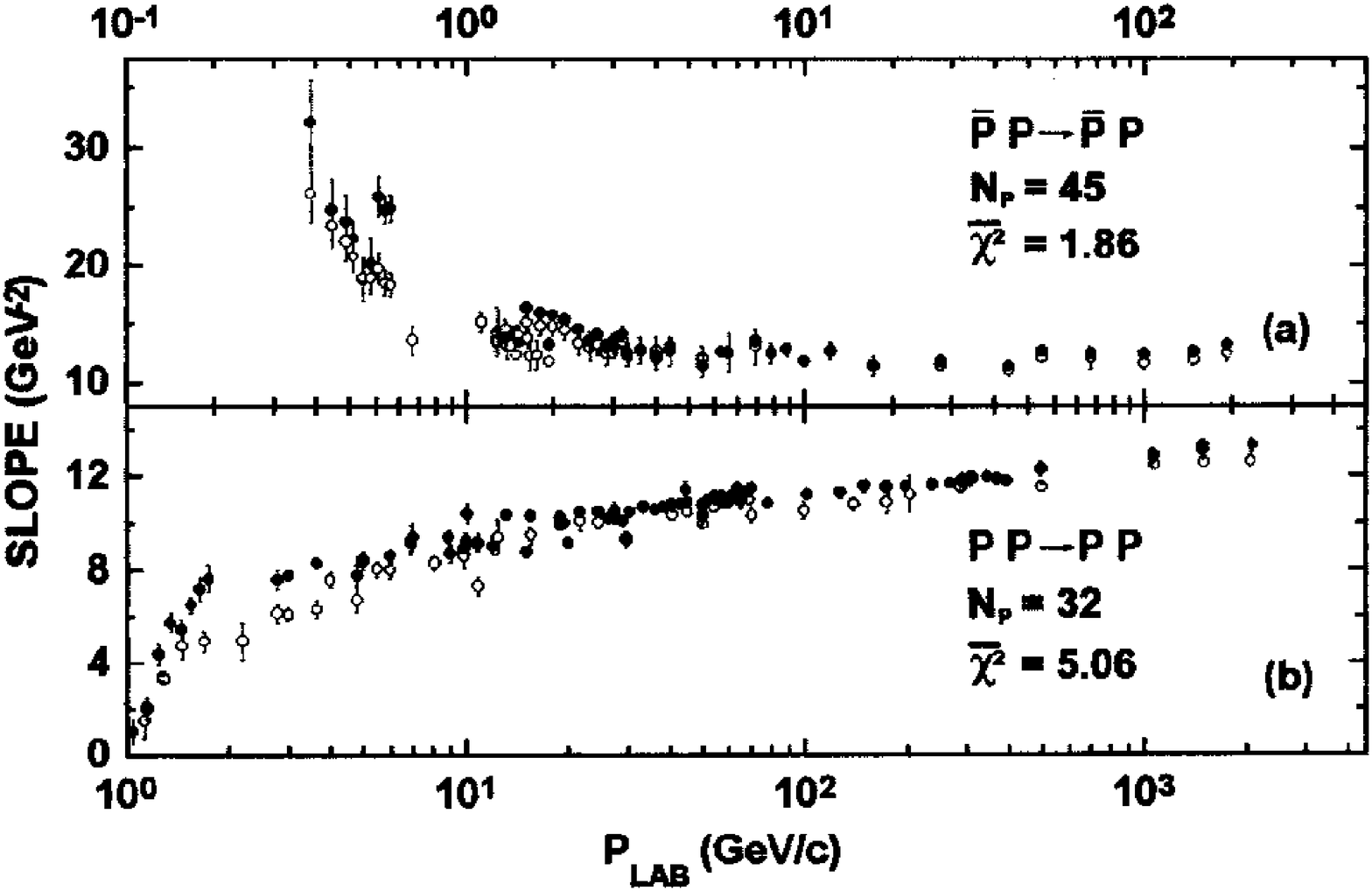, width=15.cm, height=10.cm}}\\
Fig. 2:  The experimental values (black circles) of the
logarithmic slope $b$ for the proton-proton and antiproton-proton
scatterings are compared with the {\it optimal
PMD-SS}-{\it predictions} $b_{o}$ (white
circles). The experimental data for $b$, $\frac{d\sigma }{d\Omega }%
(1)$ and $\sigma _{el}$, are taken from Refs. [24]-[26] (see the tex).
\newpage
{\par\centering \epsfig{file=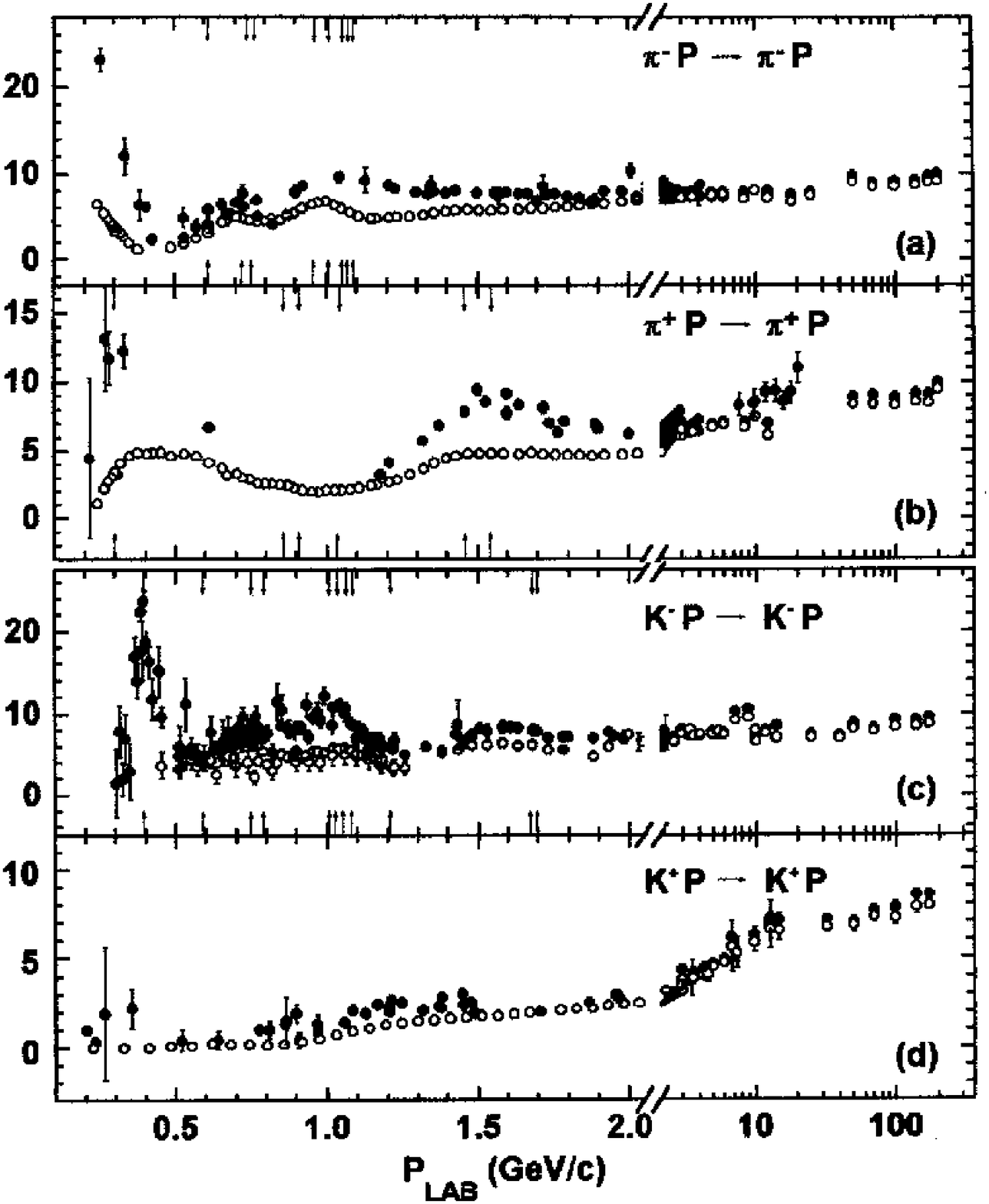, width=13.cm, height=16.cm}}\\
Fig. 3:  The experimental values (black circles) of the
logarithmic slope b for the principal meson-nucleon scatterings
are compared with the {\it optimal PMD-SS}-{\it predictions}
$b_{o}$ (white circles). The
experimental data for $b${\bf , }$\frac{d\sigma }{d\Omega }(1)$ and $%
\sigma _{el}$, are taken from Refs. [15]-[23]. The solid curves shows the
values of $b$ calculated from the experimental PSA [20,23] (see
the text). \quad \quad

\end{document}